\documentclass[%
	reprint,
	amsmath,amssymb,
	aps
]{revtex4-1}

\usepackage{graphicx}	
\usepackage{bm}			
\usepackage[colorlinks=true, linkcolor=blue,citecolor=blue]{hyperref}

\newcommand{\dd}{\text{d}}
\newcommand{\dt}{\text{dt}}
\newcommand{\dx}{\text{dx}}

\allowdisplaybreaks 

\begin{document}

\preprint{APS/123-QED}

\title{Adiabatic regularization of power spectrum in\\nonminimally coupled general single-field inflation}

\author{Allan L. Alinea}
\email{alalinea@up.edu.ph}
\altaffiliation[Corresponding Author]{}
\affiliation{
	Institute of Mathematical Sciences and Physics, University of the Philippines Los Ba\~nos, Laguna, 4031, Philippines
}
\affiliation{Department of Physics, Osaka University, Toyonaka, Osaka 560-0043, Japan}

\author{Takahiro Kubota}
\email{kubota@celas.osaka-u.ac.jp}
\affiliation{Department of Physics, Osaka University, Toyonaka, Osaka 560-0043, Japan}
\affiliation{CELAS, Osaka University, Toyonaka, Osaka 560-0043, Japan}
\affiliation{Kavli IPMU (WPI), The University of Tokyo, 5-1-5 Kashiwa-no-Ha, Kashiwa City, Chiba 277-8583, Japan}

\date{\today}

\begin{abstract}
We perform adiabatic regularization of power spectrum in nonminimally coupled general single-field inflation with varying speed of sound. The subtraction is performed within the framework of earlier study by Urakawa and Starobinsky dealing with the canonical inflation. Inspired by Fakir and Unruh's model on nonminimally coupled chaotic inflation, we find upon imposing near scale-invariant condition, that the subtraction term exponentially decays with the number of $ e $-folds. As in the result for the canonical inflation, the regularized power spectrum tends to the ``bare'' power spectrum as the Universe expands during (and even after) inflation. This work justifies the use of the ``bare'' power spectrum in standard calculation in the most general context of slow-roll single-field inflation involving non-minimal coupling and varying speed of sound.

\begin{description}
\item[PACS numbers]{98.80.Qc, 98.80.Bp, 11.10.Gh}
\end{description}
\end{abstract}

\pacs{98.80.Cq, 98.80.Qc, 11.10.Gh}	
\keywords{quantum field theory, cosmological perturbation theory, inflation, CMB}

\maketitle


\section{\label{intro}Introduction}

Cosmic inflation \cite{Guth:1980zm, Starobinsky:1980te, Starobinsky:1979ty, Sato:1980yn, Linde:1983gd, Albrecht:1982wi}, a theory involving a short period of rapid exponential expansion of space, is nowadays considered as an integral part of modern cosmology. It offers a simple solution to the horizon and flatness problems---challenges that had been hardly addressed through the Standard Big Bang cosmology before inflation was developed. In addition to this, it explains the origin of primordial density perturbations that gave rise to what we nowadays observe as galaxies and clusters of galaxies \cite{Guth:1982ec,Hawking:1982cz,Linde:1982kx,Starobinsky:1982ee}.

One of the most important physical observables in inflationary cosmology is the power spectrum of primordial density perturbations. In fact, for any viable theory of inflation, before non-Gaussianity/trispectrum \cite{Maldacena:2002vr, Bartolo:2004if, Chen:2006nt, Seery:2005wm, Alinea:2016glw, Kundu:2013gha}, tensor-to-scalar ratio \cite{Mukhanov:2005bu,Choudhury:2013iaa,Choudhury:2014sua}, loop corrections \cite{Senatore:2009cf,Seery:2007we}, and spectral index and its running \cite{Chung:2003iu,Gong:2014kpa}---quantities that have occupied a significant part of the current millennium's researches in inflationary cosmology---the power spectrum in its most basic form has to be calculated first. As such, the need for a logically consistent expression for the power spectrum based on solid physical and mathematical grounds,  cannot be overemphasized. 

In this work, we deal with one aspect of this need, by studying the effect of a subtraction procedure called \textit{adiabatic regularization} \cite{Parker:Toms, BirrellDavies, Parker:1974qw, Fulling:1974pu, Bunch:1980vc} (see also Ref. \cite{Markkanen:2013nwa} for discussion on loop correction), on the power spectrum. This is the third in a series of studies on regularizing the power spectrum following the method laid down by Urakawa and Starobinsky \cite{Urakawa:2009} within the framework of the canonical inflation model \cite{Dodelson:2003ft, LiddlenLyth, Mukhanov:book, Weinberg:2008}. The first one \cite{Alinea:2015pza} involves \textit{minimally} coupled general single-field inflation (where the speed of sound is in general, \textit{non-constant}) \cite{ArmendarizPicon:1999rj, Garriga:1999vw}, while the second one \cite{Alinea:2016qlf} deals with \textit{nonminimally} coupled chaotic inflation (where the speed of sound is \textit{constant}) \cite{Fakir:1990eg, Futamase:1987ua, Makino:1991sg, Komatsu:1999mt, Nozari:2010uu}. These two studies serve to generalize the result of Urakawa and Starobinsky in two different directions. The current work deals with \textit{nonminimally} coupled general single-field inflation where the speed of sound is in general, \textit{nonconstant} \cite{Qiu:2011, Faraoni:1996rf}. It encompasses the first two studies with the combined two layers of complexity, and hopes to extend the validity of the result in the canonical case to the general inflation model we consider here. To put this study in context, we briefly discuss below the emergence of the issue of adiabatic regularization of the power spectrum.

About a decade ago, a proposal was put forward \cite{Parker:2007} leading to a possible significant modification of the power spectrum used in standard calculation, by incorporating a subtraction term within the framework of \textit{adiabatic regularization}. To briefly elaborate the basic underlying ideas of this proposal using the labels and terminologies of the current paper, we consider the two-point function of the gauge-invariant scalar perturbation $ \mathcal R $ \cite{Bardeen:1980kt} given by
\begin{align}
	\langle
		\mathcal R(\tau , \mathbf x)\mathcal R(\tau , \mathbf y)
	\rangle
	=
	\int \frac{\dd k}{k}
	\frac{\sin (k|\mathbf x - \mathbf y|)}{k|\mathbf x - \mathbf y|} 
	\Delta ^2_{\mathcal R}(k,\tau ),
\end{align}
where $ k $ is the wavenumber, $ \mathcal R_k $ is the perturbation written in $ k $-space, $ \tau $ is the conformal time, $ (\mathbf x, \mathbf y) $ are pairs of position vectors, and $ \Delta ^2_{\mathcal R} $ is the dimensionless power spectrum defined in terms of the perturbations in $ k $-space as
\begin{align}
	\Delta ^2_{\mathcal R}(k,\tau )
	=
	\frac{k^3}{2\pi ^2}\big|\mathcal R_k(\tau )\big|^2.
\end{align}
For large values of $ k $, the Mukhanov-Sasaki equation \cite{Mukhanov:book, LiddlenLyth, Dodelson:2003ft} governing the behavior of $ \mathcal R_k, $ tells us that $ \big|\mathcal R_k(\tau )\big|^2 \sim 1/k$. It follows that in the coincidence limit $ \mathbf x \rightarrow \mathbf y $, the two-point function $ \langle	\mathcal R(\tau , \mathbf x)\mathcal R(\tau , \mathbf y) \rangle $ is quadratically divergent. 

This divergence can be removed by performing adiabatic regularization. Rather informally, the quantity $ \mathcal R_k $ on the right hand side of the first equation above is expanded in accord with the so-called \textit{adiabatic condition} \cite{Parker:Toms} and the divergent-yielding terms from this expansion are subtracted out in the integrand; thus, producing a finite result. Consistency-wise, such a regularization procedure is correspondingly reflected in the power spectrum.  Consequently, one may expect that its value can be modified after the subtraction is performed. As reported in the proposal \cite{Parker:2007}, while preserving its scale-invariant nature, the resulting physical power spectrum in the canonical inflation model differs by several orders of magnitude from the original one.

In the past decade, there have been a significant number of research papers discussing this matter about the possible modification of the power spectrum and the applicability of adiabatic regularization in the calculation of the physical power spectrum (See for instance, Refs. \cite{Parker:2007, Agullo:2008ka, Agullo:2009vq, Agullo:2009zza, Agullo:2009zi, Agullo:2010ui, Agullo:2010hg, Agullo:2011qg, Bastero-Gil:2013, Durrer:2009ii, Finelli:2007fr, Marozzi:2011da} and in particular, Ref. \cite{Bastero-Gil:2013} for a short review. Here, owing to limitation in space and a slightly different focus of this work, we are only going to outline some of the main points.) In Ref. \cite{Durrer:2009ii}, the authors argued that ``in the far infrared regime, the adiabatic expansion is no longer valid, and the unrenormalized spectra are the physical, measurable quantities.'' In Ref. \cite{Finelli:2007fr}, the authors showed that the power spectrum can become negative when one goes beyond using only those adiabatic subtraction terms sufficient to remove the divergences---the minimal subtraction scheme as we call it in this work---in the corresponding two-point function. In Ref. \cite{Bastero-Gil:2013}, the authors from a practical perspective considered the CMB and argued that in the two-point function of cosmological perturbations, only the $ \mathbf x \ne \mathbf y$ case should be considered; hence, the divergence is avoided and regularization is not necessary.

Urakawa and Starobinsky \cite{Urakawa:2009} in their work on adiabatic regularization of power spectrum in the canonical (single-field) inflation essentially reached the same conclusion that the physical power spectrum is the ``bare'' power spectrum, using a different approach. The idea is that one may perform adiabatic regularization by following the same procedure as in the original proposal \cite{Parker:2007} except for one main difference. Following our notation in this work, in the equation for regularized power spectrum given as 
\begin{align}
	\Delta^{2(r)}_{\mathcal R}
	=
	\Delta^{2(b)}_{\mathcal R}
	-
	\Delta^{2(s)}_{\mathcal R},
\end{align}
where $\Delta^{2(b)}_{\mathcal R}$ is the ``bare'' (or the original) power spectrum and $\Delta^{2(s)}_{\mathcal R}$ is the subtraction term, the authors of the original proposal evaluated the subtraction term $\Delta^{2(s)}_{\mathcal R}$ \textit{at} the horizon crossing while in Ref. \cite{Urakawa:2009}, the authors followed the evolution of $\Delta^{2(s)}_{\mathcal R}$ \textit{beyond} the horizon crossing. The latter authors found that the subtraction term decays with the number of $ e $-folds and the regularized power spectrum converges to $\Delta^{2(b)}_{\mathcal R}$.
 
In a sense, the result of Urakawa and Starobinsky is complementary to the claims in Refs. \cite{Durrer:2009ii,Marozzi:2011da,Finelli:2007fr,Bastero-Gil:2013}. They differ however, in the approach of using adiabatic regularization---it may be necessary but the final result (at least for the canonical inflation) is simply the ``bare'' power spectrum. In our past works \cite{Alinea:2015pza,Alinea:2016qlf} involving adiabatic regularization of power spectra for more general inflation models, we have followed the same scheme for three main reasons. First, \textit{consistency} of the regularization of the two-point function of primordial cosmological perturbations calls for its application to the power spectrum; that is, not only to those short-wavelength modes but to all modes the integral/summation with respect to which constitute $ \langle	\mathcal R(\tau , \mathbf x)\mathcal R(\tau , \mathbf y) \rangle $. Otherwise, some ``fundamental properties would be violated'' (e.g. the divergence of ``the renormalized energy-momentum tensor would not vanish'') \cite{Parker:2007}. However, considering the modes constituting the standard expression for the power spectrum, one should go beyond horizon crossing and follow the evolution of the subtraction term. (Note that in Ref. \cite{Durrer:2009ii}, the authors also emphasized the importance of time in applying adiabatic regularization. They offer a somewhat different perspective that adiabatic expansion is no longer valid in the superhorizon limit.) Second, we see that the coincidence limit $ \mathbf x \rightarrow \mathbf y $ is a mathematical and physical possibility that cannot be avoided by considering few practical cases of interest where $ \mathbf x $ might not be equal to $ \mathbf y$. Furthermore, one has to necessarily deal with the ``coincidence'' limit when calculating loop corrections in more advanced treatments. Third and last, the adiabatic regularization should follow the minimal subtraction prescription since in general, the adiabatic expansion is ``only asymptotic but not convergent'' \cite{Parker:Toms} and going beyond the minimal prescription can lead to unphysical negative power spectrum \cite{Finelli:2007fr}. 

In the current work, we follow the same track laid down by Urakawa and Starobinsky. As already stated above, this study encompasses the canonical case and its generalizations in two different directions covered in our previous works \cite{Alinea:2015pza, Alinea:2016qlf}. Considering the complications (i.e., nonconstant speed of sound and non-minimal coupling) brought about by our generalization, we wish to confirm the null effect of adiabatic regularization on the power spectrum.

This paper is divided as follows. In the following section, \textbf{Sec. \ref{setUp}},  we lay down the background equations from the action for our model of inflation. Then in \textbf{Sec. \ref{barePower}}, we decompose the action with respect to the primordial cosmological perturbations and then solve the ``bare'' power spectrum. The complexity due to the non-minimal coupling involved in the inflation model is addressed in this part by method of frame transformation from the Jordan frame to the Einstein frame and vice-versa. After deriving the ``bare'' power spectrum, we insert a short subsection dealing with the condition of scale invariance as it is imposed on the speed of sound. We need to impose some constraint on the speed of sound to determine the behavior of the subtraction term in which (as we shall see,) it is a part. In \textbf{Sec. \ref{adiabaticReg}}, we formally derive the subtraction term and finally perform adiabatic regularization. In the final section, \textbf{Sec. \ref{conclude}}, we state our conclusion and leave some words about our future research prospects.

\section{\label{setUp}Set up: Nonminimally Coupled General Single-Field Inflation}

The action for nonminimally coupled general single-field inflation involving the inflaton field $ \phi $ can be written as 
\begin{align}
	\label{mainAction}
	S
	=
	\frac{1}{2} \int \sqrt{-g}\,\dd^4x 
	\big[
		M_\text{Pl}^2f(\phi )R
		+
		2P(\phi,X) 
	\big],
\end{align}
where $ g $ is the determinant of the metric $ g_{\mu \nu } $, the quantity $ M_\text{Pl}^2 $ is the (square of) Planck mass, and $ R $ is the Ricci scalar. The non-minimal coupling term is included in the product $ f(\phi )R $ where $ f(\phi ) \equiv 1 + h(\phi )$. The last term on the right hand side is the ``pressure'' functional $ P $ involving $ \phi $ and the kinetic term $ X $ defined as
\begin{align}
	X
	\equiv
	-\frac{1}{2} g_{\mu \nu }
	\nabla _\mu \phi \nabla _\nu \phi.
\end{align}
In the limit where $ P \rightarrow X - V$ and $ h\rightarrow 0 $, the action above reduces to that of the slow-roll canonical inflation. 

The background spacetime for the action (\ref{mainAction}) is described by Friedmann-Lema\^itre-Robertson-Walker (FLRW) metric given by
\begin{align}
	\dd s^2 = -\dt^2 + a^2(t)\delta _{ij}\dx^i\dx^j,
\end{align}
where $ a $ is the scale factor and $ t $ is the coordinate time related to the conformal time by the definition $ \dd \tau \equiv \dt/a $. With this metric at hand and the action stated above, the equation of motion for $ \phi $ can be written as
\begin{align}
	&M_\text{Pl}^2h_\phi R
	+
	2P_\phi 
	+
	2P_{XX}\nabla _\mu X \nabla ^\mu \phi 
	\nonumber
	\\
	&\qquad
	-\,
	4P_{X\phi } X
	+
	2P_X\square\phi 
	=
	0,
\end{align} 
where the subscripts indicate partial differentiation; e.g., $ P_\phi = \partial P/\partial \phi $. Furthermore, the variation of the action with respect to $ g^{\mu \nu } $ allows us to derive the equation for the conserved energy-momentum tensor. From this tensor, we identify the energy density $ \rho $ and pressure $ p $ as
\begin{align}
	\label{energyPressure}
	\rho 
	&=
	\frac{1}{f}\big(
		P_X \dot \phi ^2 - P - 3\dot h H
	\big),
	\\[0.5em]
	p
	&=
	\frac{1}{f}(P + \ddot h + 2\dot h H). 
\end{align}
Note that our identification above for $ p $ does not lead to the identification of $ P $ as pressure as in the minimally coupled case ($h = 0$). We will however, continue to refer to $ P $ as the ``pressure'' functional.

The sum of energy density and pressure is related to the slow-roll parameter $ \epsilon \equiv -\dot H/H^2 $ through the equation
\begin{align}
	\dot H
	=
	-\frac{1}{2} (\rho + p).
\end{align}
Using the two equations above for $ \rho  $ and $ p $ we find
\begin{align}
	\epsilon
	&=
	\frac{\tilde \epsilon}{f}
	-
	\beta_1 	
	+
	\alpha 
\end{align}
where the slow-roll parameters $ \tilde \epsilon,\, \beta_1 $, and $ \alpha $ are defined as (a somewhat related quantity $ \beta _2 $ is introduced in the next section.)
\begin{align}
	\label{epsilonBeta1Alpha}
	\tilde \epsilon 
	\equiv
	\frac{\dot \phi ^2}{2H^2}P_X,
	\quad
	\beta_1
	\equiv
	\frac{\dot h}{2fH},  
	\quad  \text{and}
	\quad		
	\alpha 
	\equiv
	\frac{\ddot h}{2fH^2}.
\end{align}
The quantity $ \tilde \epsilon $ is the first slow-roll parameter in the minimally coupled case (i.e., when $ h\rightarrow 0,\,\epsilon \rightarrow \tilde \epsilon $) while $ \beta _1 $ is the same slow-roll parameter introduced in Ref. \cite{Komatsu:1999mt}. We further introduce the notation $ \alpha $ for brevity. For slow-roll inflation to take place, $ \epsilon  \ll 1 $, corresponding to the conditions $ \tilde \epsilon, |\beta |, |\alpha | \ll 1$.

In this work, as in the canonical case, we assume slow-roll inflation. We add a further limitation that the type of inflation be chaotic. Consequently, we may impose that it start out with large value of $ h(\phi) $ and ends when $ h(\phi ) \sim \mathcal O(1) $. This is in accord with the nonminimally coupled chaotic inflation considered in Refs. \cite{Fakir:1990eg, Futamase:1987ua, Makino:1991sg, Komatsu:1999mt}   where $ h(\phi ) = \xi \phi ^2/M_\text{Pl}^2 $ with $ \xi  $ as the non-minimal coupling parameter usually taken to be much greater than unity \cite{Fakir:1990eg, Salopek:1988qh}. Considering the power spectrum of the primordial cosmological perturbations, all our calculations here are good at least to first order in the slow-roll parameters. As far as adiabatic regularization of the power spectrum is concerned, this should be more than sufficient. Lastly, we do not consider here loop corrections.

\section{\label{barePower}The ``Bare'' Power Spectrum, Scale Invariance, and Speed of Sound}
\subsection{``Bare'' Power Spectrum}
After setting up the necessary background equations in the previous section, we now turn our attention to fluctuation. The perturbed metric containing the fluctuation $ \mathcal R $ can be written in the ADM decomposition \cite{Arnowitt:1959ah} as
\begin{align}
	\dd s^2
	=
	-N^2\dt^2
	+
	g_{ij}(\dx^i + N^i\dt)(\dx^j + N^j\dt),
\end{align}
where $ N $ and $ N^j $ are the \textit{lapse} and \textit{shift} functions respectively. We wish to calculate the ``bare'' power spectrum of $ \mathcal R $ (in $ k$-space) using the \textit{comoving} gauge \cite{Maldacena:2002vr} where $ \delta \phi = 0 $, and the spatial part of the metric above takes the form
\begin{align}
	g_{ij}
	=
	a^2(t)e^{2\mathcal R}\delta _{ij}\dx^i\dx^j.
\end{align}

In the minimally coupled model of inflation, the (``bare'') power spectrum can be computed by first decomposing the action $S_\text{m}$ given by 
\begin{align}
	\label{minAction}
	S_\text{m}
	=
	\frac{1}{2} \int \sqrt{-g}\,\dd^4x 
	\big[
		M_\text{Pl}^2R
		+
		2P(\phi,X) 
	\big],
\end{align}
with respect to $ \mathcal R $ as $ S_\text{m} = S_\text{m}^{(0)} + S_\text{m}^{(2)} + S_\text{m}^{(3)} + \cdots$, subject to the chosen gauge and the metric decomposition above. (For calculations in the minimally coupled case, interested readers may see Refs. \cite{Maldacena:2002vr,Chen:2006nt}.) Then the equation of motion for $ \mathcal R $ called the \textit{Mukhanov-Sasaki equation}, is derived from the second order action $ S_\text{m}^{(2)} $. In $ k $-space, it states
\begin{align}
	v_k'' + \bigg(k^2c_s^2 - \frac{z''}{z} \bigg)v_k
	&=
	0,
\end{align}
where the speed of sound $ c_s^2, $ perturbation $ v_k, $ and $ z $ in the potential term $ z''/z $, are given by
\begin{align}
	v_k \equiv z\mathcal R_k,
	\quad
	z^2 \equiv \frac{2a^2\epsilon}{c_s^2},
	\quad
	c_s^2
	=
	\frac{P_X}{P_X + 2XP_{XX}}, 
\end{align}
and the symbol prime indicates differentiation with respect to $ \tau $. Finally, the solution of Mukhanov-Sasaki equation is computed and the ``bare'' power spectrum is determined using the equation 
\begin{align}
	\label{barePowSpec}
	\Delta ^{2(b)}_{\mathcal R}
	=
	\frac{k^3}{2\pi ^2} \big|
		\mathcal R_k
	\big|^2.
\end{align}
Note that implicit in this procedure is the quantization of perturbation and calculation of the vacuum expectation value with respect to the Bunch-Davies vacuum.

In the nonminimally coupled model of inflation that we consider in this work, one may follow the same procedure. However, the decomposition of the action can become quite complicated because of the presence of the non-minimal coupling term $ h(\phi )R $. To ease our way of establishing the Mukhanov-Sasaki equation, we use the method of frame transformation from the Jordan frame (where the original model is in place) to the Einstein frame; then we go back to the Jordan frame with respect to which we write down the power spectrum. In using this method, the metric is transformed as
\begin{align}
	\dd s^2 \rightarrow \dd \widehat s\,^2 = \Omega ^2(\phi ) \dd s^2,
\end{align}
where the hat indicates Einstein frame and $ \Omega^2 $ is called the \textit{conformal factor} that we choose as $ \Omega ^2 = f(\phi ) $. The main advantage of transforming to the Einstein frame is that in this frame, the action takes a minimally coupled form. As such, the relations in this frame are the same as those of minimally coupled case but with the variables involved ``wearing'' a hat. Furthermore, noting that the perturbation is frame invariant (i.e., $ \widehat {\mathcal R} = \mathcal R$) \cite{Makino:1991sg, Sugiyama:2010zz, Gong:2011qe, Chiba:2008ia} one may perform the decomposition $ \widehat S = \widehat S^{(0)} + \widehat S^{(2)} + \widehat S^{(3)} + \cdots$ \cite{Kubota:2011re} and from $ \widehat S^{(2)} $ establish the Mukhanov-Sasaki equation as
\begin{align}
	\label{MSEqEinstein}
	\widehat v_k\hspace{-0.23em}'' + \bigg(k^2\widehat c_s\hspace{-0.23em}^2 
	- 
	\frac{\widehat z\,''}{\widehat z} \bigg)\widehat v_k
	&=
	0,
\end{align}
where 
\begin{align}
	\label{vCsXhZh}
	\widehat v_k = \widehat z\,\mathcal R_k,
	\;
	\widehat c_s\hspace{-0.23em}^2
	=
	\frac{\widehat P_{\widehat X}}{
		\widehat P_{\widehat X}
		+
		2\widehat X \widehat P_{\widehat X \widehat X}
	},
	\;
	\widehat X
	=
	\frac{X}{\Omega ^2},
	\;
	\widehat z\,^2
	=
	\frac{2\widehat a^2 \widehat \epsilon }{\widehat c_s\hspace{-0.23em}^2},   
\end{align}
with
\begin{align}
	&\widehat P
	=
	\frac{P}{\Omega ^4}
	+
	6M_\text{Pl}^2 \widehat X
	\frac{\Omega _\phi ^2}{\Omega ^2},
	\quad 
	\widehat a
	\equiv
	\Omega a,
	\quad
	\widehat \epsilon 
	\equiv
	-\frac{1}{\widehat H^2}
	\frac{\dd \widehat H}{\dd\widehat t},
	\nonumber
	\\
	&\widehat H
	\equiv
	\frac{\dd\widehat a}{\dd\widehat t},
	\quad
	\text{and}
	\quad
	\dd\widehat t
	\equiv
	\Omega \dt.    
\end{align}
Needless to say, the conformal time is invariant under conformal transformation (i.e., $ \widehat \tau = \tau $) so there is no need to have another notation for the derivative in the Mukhanov-Sasaki equation above.

It then remains for us to solve (or more precisely, semi-analytically approximate the solution of) the Mukhanov-Sasaki equation (\ref{MSEqEinstein}) to find the ``bare'' power spectrum. To do this, we need to express the potential term $ \widehat z\,''/\widehat z $ in terms of the conformal time and corrections due to slow-roll parameters. The quantity $ \widehat z $ depends on $ \widehat \epsilon,\, \widehat c_s^{\,2} $, and $ \widehat a \equiv \Omega a$. For the hatted first slow-roll parameter, we find upon using its definition given above, 
\begin{align}
	\label{epsilonHat}
	\widehat \epsilon 
	=
	\frac{\tilde \epsilon}{\Omega ^2}\frac{p_2}{p_1^2},
\end{align}
where
\begin{align}
	p_1 
	\equiv
	1 + \beta _1,
	\;
	p_2 
	\equiv 
	1 
	+ 
	s_2,
	\;
	\text{with}
	\;
	s_2
	\equiv
	\frac{3M_\text{Pl}^2}{2}   
	\frac{h_\phi ^2}{\Omega ^2P_X}.
\end{align}
Note that unlike $ |\beta _1| \ll 1 $, the quantity $ s_2 $ is not necessarily small during inflation. In the Fakir and Unruh model of nonminimally coupled chaotic inflation \cite{Fakir:1990eg} for instance, $ s_2 $ is nearly constant and of the order  $10^3$ during inflation for $ \xi \sim \mathcal O(10^3) $. Here, we assume this same behavior and take the magnitude of 
\begin{align}
	\label{beta2Define}
	\beta _2 \equiv \frac{\dot p_2}{p_2 H} 
\end{align} 
to be much less than unity. For the hatted speed of sound, we find from the second of (\ref{vCsXhZh})
\begin{align}
	\label{csHat}
	\widehat c_s^{\,-2}
	=
	c_s^{-2} 
	\frac{1 + c_s^2\, s_2}{1 + s_2}
	=
	c_s^{-2}(1 + \kappa _s),
\end{align}
where $ 1 + \kappa _s \equiv (1 + c_s^2\,s_2)/(1 + s_2) $. We assume in this work that $ |\kappa _s| \ll 1 $ and is of the same order of magnitude as that of the slow-roll parameters. 

Using the results above for $ \widehat \epsilon $ and $ \widehat c_s^{\,-2} $ and the definition $ \widehat a \equiv \Omega a $, in the expression for $ \widehat z $ given by the last of (\ref{vCsXhZh}) we find
\begin{align}
	\label{zHatSq}
	\widehat z^{\,2}
	&=
	\frac{2 a^{2}\tilde \epsilon}{c_s^2}\frac{ p_2}{p_1^2} (1 + \kappa _s)\,.
\end{align}
It is then straightforward to determine the potential term from this expression. To first order in the slow-roll parameters we have
\begin{align}
	\frac{\widehat z''}{\widehat z} 
	&=
	(aH)^2 \big(
		2 - \epsilon + \tfrac{3}{2}\,\tilde \epsilon _2
		+
		3\delta _1
		+
		\tfrac{3}{2}\,\beta _2	
		+
		\cdots 	
	\big),
\end{align}
where $ \tilde  \epsilon _2 \equiv \dot {\tilde {\epsilon }}/\tilde \epsilon H $ is the \textit{second Hubble flow parameter} \cite{Schwarz:2001vv} in the minimal coupling limit and $ \delta _1 \equiv -\dot c_s/c_sH $ is the \textit{first sound flow-parameter} \cite{Martin:2013uma}. (Note that the usual slow-roll parameters are $ \epsilon \equiv -\dot H/H^2$ and $\eta \equiv \epsilon - \dot \epsilon /2\epsilon H$. In terms of the Hubble flow parameters $\{\epsilon _i\}$, they can be written as $ \epsilon = \epsilon _1 $ so we simply write $ \epsilon _1 $ as $ \epsilon $, and $ \eta = \epsilon - \epsilon _2/2 $. As always, the tilde corresponds to the minimal coupling limit.) The factor $ aH $ on the right hand side of the equation for the potential term is related to the conformal time as \cite{Alinea:2015gpa}
\begin{align}
	\tau 
	&=
	-\frac{1}{a}\sum_{n = 0}^\infty
	\bigg(
		H^{-1}\frac{\dd}{\dt} 
	\bigg)^n
	H^{-1},
	\nonumber
	\\
	\tau 
	&=
	-(aH)^{-1}\big(1 + \epsilon + \cdots\big).
\end{align} 
This allows us to finally write $ \widehat z\,''/\widehat z $ in terms of $ \tau  $ as
\begin{align}
	\label{zHatPP}
	\frac{\widehat z''}{\widehat z} 
	&=
	\frac{1}{\tau ^2} \big(
		2 + 3\epsilon + \tfrac{3}{2}\,\tilde  \epsilon _2
		+
		3\delta _1
		+
		\tfrac{3}{2}\,\beta _2		
		+
		\cdots 	
	\big).
\end{align}

If the hatted speed of sound is unity, with the potential given above, the Mukhanov-Sasaki equation (\ref{MSEqEinstein}) can be readily transformed to Bessel differential equation and be solved through the Hankel function approximation. The presence of $ \widehat c_s^{\,2} $ in the term $ k^2\widehat c_s^{\,2} $ complicates the differential equation because of its dependence on $ \tau  $. To remedy this, we define a new independent variable $ \widehat y $ involving $ \widehat c_s $ and $ \tau  $ and rewrite the Mukhanov-Sasaki equation as
\begin{align}
	\label{newMukhanovSasakiEqn}
	\frac{\dd^2\widehat u_k}{\dd\widehat y^2} 
	+
	\bigg(
		k^2
		-
		\frac{1}{\widehat q} 
		\frac{\dd^2\widehat q}{\dd\widehat y^2} 
	\bigg)\widehat u_k
	=
	0,
\end{align}
where
\begin{align}
	\widehat q
	\equiv \widehat z \sqrt{\widehat c_s},
	\quad
	\widehat u_k
	\equiv
	\widehat v_k\sqrt{\widehat c_s},
	\quad
	\text{and}
	\quad
	\dd\widehat y
	\equiv
	\widehat c_s\,\dd \tau.   
\end{align}
We can explicitly write $ \widehat y $ in terms of $ c_s $ and $ \tau  $ as
\begin{align}
	\widehat y
	=
	c_s\tau \big(
		1 - \delta _1 - \tfrac{1}{2}\kappa _s
	\big).
\end{align}
Moreover, the new potential term can be expressed in terms of the old potential term and derivatives of $ \widehat c_s $ through the definitions above for $ \widehat q $ and $ \widehat y $. This can then be manipulated to gain an equation analogous to (\ref{zHatPP}) for $ \widehat z''/\widehat z $:
\begin{align}
	\frac{1}{\widehat q} 
	\frac{\dd^2\widehat q}{\dd\widehat y^2}	
	&=
	\frac{1}{\widehat y^{\,2}}
	(2 + 3 \epsilon_{ps} + \tfrac{3}{2}\,\tilde  \epsilon _2 ),
\end{align}
where we have defined $ \epsilon _{ps} \equiv \epsilon - \frac{1}{2}\delta _1 + \frac{1}{2} \beta _2$. Note the absence of $ \kappa _s $ in the set of small parameters inside the pair of parentheses on the right hand side. It is present in the intermediate calculation involving the relationship between $ \widehat c_s $ and $ c_s $ but cancels along the way.

With the above result for the new potential term, we can rewrite (\ref{newMukhanovSasakiEqn}) as
\begin{align}
	\frac{\dd^2\widehat u_k}{\dd\widehat y^2}
	+
	\bigg[
		k^2
		-
		\frac{1}{\widehat y^{\,2}}\bigg(
			\nu _s^2 - \frac{1}{4} 
		\bigg) 
	\bigg]\widehat u_k
	=
	0
\end{align}
with $ 	\nu _s = \tfrac{3}{2} +	\epsilon _{ps} + \tfrac{1}{2}\,\tilde \epsilon _2 $. Defining $ \widehat x = -k\widehat y $ and $ \widehat u_k = \widehat w_k \sqrt{\widehat x}  $ allows us to transform this to Bessel differential equation,
\begin{align}
	\widehat x^{\,2}
	\frac{\dd^2\widehat w_k}{\dd\widehat x^{\,2}} 
	+
	\widehat x\frac{\dd\widehat w_k}{\dd\widehat x}
	+
	\big(
		\widehat x^{\,2} - \nu _s^2
	\big) \widehat w_k
	=
	0,
\end{align}
the solution of which can be written in terms of the Hankel functions. Going back to $ \widehat u_k $ and then to $ \widehat v_k $, we find in the superhorizon limit $ kc_s \ll aH $ that
\begin{align}
	\widehat v_k
	&=
	\frac{1}{2} \sqrt{\frac{\pi }{k} } 
	e^{i\frac{\pi }{4} \left(2\nu _s + 1\right)}	
	\sqrt{\frac{\widehat x}{\widehat c_s} } 
	\bigg[
		-\frac{i}{\pi }\Gamma (\nu _s)
		\left(\frac{\widehat x}{2} \right) ^{-\nu _s}
	\bigg],
\end{align}
where $ \Gamma $ is the gamma factorial function. 

We substitute the expression for $ \widehat v_k $ above into $ \mathcal R = \widehat v_k/\widehat z_k $ and then to the equation for the ``bare'' power spectrum given by (\ref{barePowSpec}). After performing some algebraic manipulation, we finally obtain 
\begin{widetext}
	\begin{align}
		\label{barePowSpecIi}
		\Delta ^{2(b)}_{\mathcal R}
		=
		\frac{ H_*^2}{8\pi ^2\tilde \epsilon_* c_{s*}p_{2*}}
		\big[
			1 - 2\epsilon _{*} + 2\delta _{1*} + 2\beta _{1*} + \tfrac{1}{2}\kappa _{s*}
			+
			(2 \epsilon _{ps*} + \tilde \epsilon _{2*})
			(2 - \gamma  - \ln 2)
		\big],	
	\end{align}
\end{widetext}
where $ \gamma = 0.5772 $ is the Euler-Mascheroni constant and `*' indicates horizon crossing. This is our sought-for expression for the ``bare'' power spectrum. Observe that in the limit where $ c_{s*} \rightarrow 1$ and $ h\rightarrow 0 $, it reduces to that of the canonical case as
\begin{align}
	\Delta ^{2(b)}_{\mathcal R}
	\rightarrow
	\frac{ H_*^2}{8\pi ^2\bar \epsilon_*}
	\big[
		1 
		-
		2\bar \epsilon _{*}
		+
		(2\bar \epsilon _{*} + \bar \epsilon _{2*})
		(2 - \gamma  - \ln 2)
	\big],
	\nonumber
\end{align}
where $ \bar \epsilon = \dot \phi ^2/2H^2 $ is the canonical first slow-roll parameter and $\dot {\bar \epsilon} = H\bar \epsilon\, \bar\epsilon_2$. For non-minimally coupled chaotic inflation case where the speed of sound is constant and $ h = \xi \phi ^2/M_\text{Pl}^2  $, Eq. (\ref{barePowSpecIi}) reduces to
\begin{widetext}
	\begin{align}
		\Delta ^{2(b)}_{\mathcal R}
		\rightarrow
		\frac{ H_*^2}{8\pi ^2\bar \epsilon_* (1 + 6\xi )}
		\big[
			1 - 2\epsilon _{*} + 2\beta _* 
			+
			(2 \epsilon _{*} +\beta _{2*} + \bar \epsilon _{2*})
			(2 - \gamma  - \ln 2)
		\big],	
	\end{align}
\end{widetext}
which is in agreement with the result in Refs. \cite{,Fakir:1990eg, Makino:1991sg, Komatsu:1999mt} (apart from the slow-roll corrections not included in these references). 

Note that in Ref. \cite{,Qiu:2011}, the authors derived the expression for the ``bare'' power spectrum for essentially the same model we considered here using a different method following a somewhat ``brute-force'' decomposition of the action in the Jordan frame. Their result cannot be readily compared to ours due to a different set of slow-roll parameters used. The expression for the power spectrum also involves $ f(\phi )R $ evaluated at an unusual value of $ \tau = 1 $. In addition to this, some slow-roll parameters seem to have been perhaps, unintentionally omitted.

\subsection{Scale Invariance and the Speed of Sound}
\textit{``Exact'' Scale Invariance.} In Ref. \cite{Khoury:2008wj} (see also Refs.  \cite{ArmendarizPicon:2003ht, Magueijo:2008pm, Piao:2006ja}), the authors investigated scale invariance of the power spectrum and its relationship to the speed of sound within a minimally coupled framework with Lagrangian $ \mathcal L = \sqrt{-g}\,P(\phi ,X)  $. Part of this study was establishing a relationship between $ \delta _1 $ and $ \epsilon $ as imposed by scale invariance. They found that \textit{exact} scale invariance requires that both of these parameters be constant. Moreover, there are two possible cases realizing this namely, (a) $ \delta _1 = 2 \epsilon $ corresponding to an \textit{expanding universe with a decreasing sound speed} and (b) $ \delta _1 = -\frac{2}{5}(3 - 2 \epsilon )  $ corresponding to a \textit{contracting universe with increasing sound speed}. 

In this work, we need to impose some constraint on the speed of sound as part of the determination of the behavior of the subtraction term in the adiabatic regularization of the power spectrum (see the following section). Inspired by symmetry, we impose near scale invariance and similar to that in Ref. \cite{Khoury:2008wj}, find the relationship between $ \delta _1 $ and $ \epsilon $. Following a similar track as that of the mentioned reference, we first, consider an ideal scenario corresponding to ``exact'' scale invariance. Then, we consider deviation about this and arrive at a more realistic equation corresponding to a near scale-invariant condition on the speed of sound. 

For the choice of an expanding universe, we follow a simplified derivation using the expression for the spectral tilt (instead of starting with the Mukhanov-Sasaki equation as was done in Ref. \cite{Khoury:2008wj}). From the expression for the power spectrum given by (\ref{barePowSpecIi}) we find the spectral tilt to first order in the slow-roll parameters as
\begin{align}
	\label{diffBarePowSpec}
	\frac{\dd \ln \Delta ^{2(b)}_{\mathcal R}}{\dd\ln k} 
	&=
	\frac{\dd\ln \Delta ^{2(b)}_{\mathcal R}}{\dd N_*}
	\frac{\dd N_*}{\dd\ln k},
	\nonumber
	\\[0.5em]
	\frac{\dd \ln \Delta ^{2(b)}_{\mathcal R}}{\dd\ln k} 
	&=
	-2 \epsilon _{1*}
	+
	\delta _{1*}
	-
	\beta _{2*}
\end{align}
Exact scale invariance then dictates $ -2 \epsilon _{1*} + \delta _{1*} -	\beta _{2*} = 0,$ which is the same as that of the minimally coupled inflation scenario except for the addend $-\beta _{2*} $. 

Note however, that within the framework of Fakir and Unruh's model of nonminimally coupled chaotic inflation \cite{Fakir:1990eg} where $ h = \xi \phi ^2/M_\text{Pl}^2  $ with $ \xi \sim \mathcal O(10^3) $, the quantity $\beta_{2*} \approx 2\beta_{1*}/h \ll \beta_{1*}, $ where $\beta_1$ is one of the slow-roll parameters. Consequently, within this framework, 
\begin{align}
	\label{scaleInvariantNonMin}
	\delta _{1*} = 2 \epsilon _* + \text{higher order correction},
\end{align}
which is essentially the same as that of the minimally coupled inflation. In this work, considering the more general model that we have, we assume that the contribution of  $\beta_2$ in the scale-invariant condition is also sub-dominant.

\medskip
\textit{Near Scale Invariance.} The scale-invariance condition (\ref{scaleInvariantNonMin}) ties up the variation of the speed of sound $\delta_1$ and the behavior of the Hubble parameter as measured by $\epsilon_1$ so tightly with little freedom to accommodate a realistic inflation scenario. This condition has at least two limitations. First, the way it is written does not include the element of time. Second, in the special case where the speed of sound is constant, $\delta_1 \rightarrow 0$ and the slow-roll parameter $\epsilon$ is also forced to become too small or vanishing if the higher corrections are negligible.

To remedy these limitations, we assume a near scale-invariant condition given by 
\begin{align}
	\label{nearScaleInvariant}
	\delta _1(\tau ) = 2\epsilon(\tau ) - \alpha _s(\tau ).
\end{align}
Here, in the general case where the speed of sound is not constant, we take $\alpha_s$ as subdominant, that is, at least of the second order in the slow-roll parameters, such that $\delta_1 = 2\epsilon$ to first order. To accommodate the specific case of constant speed of sound, we assume that in the limit as $\delta_1 \rightarrow 0$, $\alpha_s \rightarrow 2\epsilon$ so that the equation becomes $\epsilon = \epsilon$ (trivial). Although it does not impose anything on $\epsilon$, we have a slow-roll assumption in place. Near scale invariance in this case hinges on the smallness of $\epsilon$ during a significant part of inflation.

Note that the near scale invariant condition given above is in agreement with Ref. \cite{Khoury:2008wj}. In this research article, the authors arrived at the equation given by (partially using our notation,) $\delta_1 = 2\epsilon - (1 + \epsilon)\delta_I$, where $\delta_I$ is a parameter that depends at least on $\epsilon$ and $\delta _1$. Analysis of the calculations in the mentioned reference indicates that this equation behaves the same way as our near scale invariance equation above with $\alpha_s = (1 + \epsilon)\delta_I$.

\section{Adiabatic Regularization of Power Spectrum}
\label{adiabaticReg}
The regularized power spectrum $ \Delta ^{2(r)}_{\mathcal R} $ is the difference between the ``bare'' power spectrum $ \Delta ^{2(b)}_{\mathcal R} $ calculated in the immediately preceding section, and the subtraction term $ \Delta ^{2(s)}_{\mathcal R} $; symbolically,
\begin{align}
	\label{regPowSpec}
	\Delta ^{2(r)}_{\mathcal R}
	=
	\Delta ^{2(b)}_{\mathcal R}
	-
	\Delta ^{2(s)}_{\mathcal R}.
\end{align}
In this section we wish to derive the form of $ \Delta ^{2(s)}_{\mathcal R} $ and perform the subtraction process. Moreover, we investigate the behavior of the regularized power spectrum as the Universe expands; that is, with respect to the number of $ e $-folds.

\subsection{Derivation of the Subtraction Term}
The subtraction term can be written in terms of $ \mathcal R^{(s)}_{k} $ (now with a superscript `s' for ``subtraction'') following the form of the ``bare'' power spectrum given by (\ref{barePowSpec}). We have
\begin{align}
	\label{subtracTermPowSpec}
	\Delta ^{2(s)}_{\mathcal R}
	=
	\frac{k^3}{2\pi ^2} 
	\big|\mathcal R^{(s)}_{k}\big|^2
	=
	\frac{k^3}{2\pi ^2} 
	\bigg|\frac{v^{(s)}_{k}}{z} \bigg|^2,
\end{align}
where $ v_k^{(s)} \equiv z\mathcal R^{(s)}_{k} $. The quantity $ v_k^{(s)} $ is given by the ansatz (see Refs. \cite{Parker:Toms,BirrellDavies})
\begin{align}
	\label{vksNoHat}
	v_k^{(s)}(\tau )
	=
	\frac{1}{\sqrt{2W_k(\tau )} }
	e^{-i\int ^\tau \dd\tilde \tau \,W_k(\tilde \tau ) },
\end{align}
that resembles the plane-wave solution of the Mukhanov-Sasaki equation. In adiabatic regularization, $ W_k(\tau ) $ that we simply write hereafter as $ W $ for brevity, is expanded as $ W = \omega _0 + \omega _1 + \omega _2 + \cdots,$ where the subscript $ n $ in $ \omega _n $ denotes the \textit{adiabatic order}. Furthermore, one imposes the \textit{adiabatic condition} on $ v_k^{(s)}(\tau ) $. Put simply, this means that $ v_k^{(s)}(\tau ) $ should reduce to the plane-wave solution in the limit of very slow expansion of the Universe or very small wavelength of the modes involved, such that spacetime is effectively flat.

Similar to that of the determination of $ \Delta ^{2(b)}_{\mathcal R} $, there is an added layer of complexity in finding $ \Delta ^{2(s)}_{\mathcal R} $ due to the presence of non-minimal coupling term. For the former, we performed frame transformations and exploited the fact that $ \widehat {\mathcal R}^{(b)}_k = \mathcal R^{(b)}_k  $ to easily find $ \mathcal R^{(b)}_k $ and in turn, $ \Delta ^{2(b)}_{\mathcal R} $. For the latter, it would be good if we could perform the same frame transformations and exploit a similar relation between $ \widehat {\mathcal R}^{(s)}_k $ and $ \mathcal R^{(s)}_k  $. As it turns out, as argued in Ref. \cite{,Alinea:2015pza}, $ \widehat {\mathcal R}^{(s)}_k = \mathcal R^{(s)}_k  $; in other words, it is also frame-invariant.

Working in the Einstein frame, following (\ref{vksNoHat}), one writes $ \widehat v_k^{(s)}(\tau )$ in terms of $ \widehat W $ and then expand the latter as $\widehat W = \widehat \omega _0 + \widehat \omega _1 + \widehat \omega _2 + \cdots$. The working equation for $ \widehat W $ can be simply borrowed from the minimally coupled case we investigated in Ref. \cite{Alinea:2015gpa}, but with all the corresponding variables involved ``wearing'' a hat:
\begin{align}
	\frac{\widehat W''}{2\widehat W} 
	-
	\frac{3}{4}
	\frac{\widehat W'^2}{\widehat W^2}
	+
	\widehat W^2
	-
	k^2\widehat c_s^{\,2}  
	+
	\frac{\widehat z''}{\widehat z}
	=
	0. 
\end{align}
Performing a substitution from the expansion of $ \widehat W $ above in the working equation, equating terms of the same adiabatic order, and considering only the expansion up to second adiabatic order based on the minimal subtraction prescription, we find 
\begin{align*}
	\Big|
		\mathcal R_k^{(s)}
	\Big|^2
	=
	\frac{1}{2\widehat z^{\,2}k\widehat c_s} \bigg[
		1
		+
		\frac{1}{2k^2 \widehat c_s^{\,2}} 
		\frac{\widehat z''}{\widehat z} 
		+
		\frac{1}{k^2 \widehat c_s^{\,2}} \bigg(
			\frac{1}{4}
			\frac{\widehat c_s\hspace{-0.23em}''}{\widehat c_s}  
			-
			\frac{3}{8}
			\frac{\widehat c_s\hspace{-0.23em}'^{2}}{\widehat c_s^{\,2}}  
		\bigg)
	\bigg].
\end{align*}
Note that for superhorizon modes, the first term inside the pair of square brackets is negligible compared to the other terms. We are then left with the task of expanding the second term involving the potential, and the derivative terms involving the hatted speed of sound.

Using the expressions for $ \widehat c_s $ and $ \widehat z $ given by (\ref{csHat}) and (\ref{zHatSq}) respectively, the equation for $ |\mathcal R^{(s)}_k(\tau )|^2 $ can be rewritten as
\begin{align}
	\Big|
		\mathcal R_k^{(s)}
	\Big|^2
	&=
	\frac{H^2}{4k^3\tilde \epsilon c_s p_2}\big(
		1 - \tfrac{1}{2}\,\epsilon 
		+
		\tfrac{3}{4}\,\tilde \epsilon _2
		+
		\tfrac{5}{4}\,\delta _1
		+
		2\beta _1
		\nonumber
		\\
		&\qquad\qquad\qquad
		+
		\tfrac{3}{4}\,\beta _2
		+
		\tfrac{1}{2}\,\kappa _s
	\big).	
\end{align}
This can be readily substituted in the equation for $ \Delta ^{2(s)}_k $ given by (\ref{subtracTermPowSpec}) to gain
\begin{align}
	\label{subtractTerm}
	\Delta ^{2(s)}_{\mathcal R}
	&=
	\frac{H^2}{8\pi ^2\tilde  \epsilon c_s p_2}\big(
		1 - \tfrac{1}{2}\,\epsilon 
		+
		\tfrac{3}{4}\,\tilde \epsilon _2
		+
		\tfrac{5}{4}\,\delta _1
		+
		2\beta _1
		\nonumber
		\\
		&\qquad\qquad\qquad
		+\,
		\tfrac{3}{4}\,\beta _2
		+
		\tfrac{1}{2}\,\kappa _s
	\big).	
\end{align}
This is our sought-for expression for the subtraction term. Observe that in the non-minimal coupling limit $ h \rightarrow 0 $ where the speed of sound is in general non-constant, $ p_2\rightarrow 1,\, \beta _1\rightarrow 0, $ and $ \kappa _s \rightarrow 0 $. In such a case, 
\begin{align}
	\Delta ^{2(s)}_{\mathcal R}
	&\rightarrow
	\frac{H^2}{8\pi ^2 \tilde \epsilon c_s }\big(
		1 - \tfrac{1}{2}\,\tilde \epsilon 
		+
		\tfrac{3}{4}\, \tilde \epsilon _2
		+
		\tfrac{5}{4}\,\delta _1
	\big),
\end{align}
consistent with the result in Ref. \cite{,Alinea:2015pza}. When we further take the limit $ c_s\rightarrow 1 $ of the above expression as in the canonical case, it reduces to the result of Urakawa and Starobinsky \cite{Urakawa:2009}, namely,
\begin{align}
	\Delta ^{2(s)}_{\mathcal R}
	&\rightarrow
	\frac{H^2}{8\pi ^2 \bar \epsilon }\big(
		1 - \tfrac{1}{2}\,\bar \epsilon 
		+
		\tfrac{3}{4}\, \bar\epsilon _2
	\big).
\end{align}

\bigskip
\bigskip
\subsection{The Regularized Power Spectrum}
Now that the ``bare'' power spectrum and the subtraction term are in place, we can finally compute the regularized power spectrum. By virtue of (\ref{regPowSpec}) for $ \Delta ^{2(r)}_{\mathcal R}, $ and (\ref{barePowSpecIi}) and (\ref{subtractTerm}) for $ \Delta ^{2(b)}_{\mathcal R} $ and $ \Delta ^{2(s)}_{\mathcal R} $ respectively, we find
\begin{widetext}
	\begin{align}
		\label{regPowSpecBoxed}
		\Delta ^{2(r)}_{\mathcal R} 
		=
		\frac{ H_*^2}{8\pi ^2\tilde \epsilon_* c_{s*}p_{2*}}\bigg[
			1 + \epsilon ^{(b)}_*
			-
			\bigg(\frac{H^2}{H_*^2}\bigg)
			\bigg(\frac{\tilde \epsilon _*}{\tilde \epsilon } \bigg)
			\bigg(\frac{p_{2*}}{p_2} \bigg)
			\bigg(\frac{c_{s*}}{c_s}\bigg)
			\big(1 + \epsilon ^{(s)}\big)
		\bigg],
	\end{align}
\end{widetext}
where we have lumped together the slow-roll corrections as
\begin{align}
	\epsilon ^{(b)}_*
	&\equiv
	-2\epsilon _{*} + 2\delta _{1*} + 2\beta _* 
	+ 
	\tfrac{1}{2}\kappa _{s*}
	\nonumber
	\\
	&\qquad
	+\,
	(2 \epsilon _{ps*} 
	+
	\tilde \epsilon _{2*})
	(2 - \gamma  - \ln 2),	
	\nonumber
	\\[0.5em]
	\epsilon ^{(s)}
	&\equiv
	- \tfrac{1}{2}\,\epsilon 
	+
	\tfrac{3}{4}\,\tilde \epsilon _2
	+
	\tfrac{5}{4}\,\delta _1
	+
	2\beta _1
	+
	\tfrac{1}{2}\,\kappa _s.
\end{align}
The first two terms $(1 + \epsilon ^{(b)}_*)$ inside the pair of square brackets in the equation above for $ \Delta ^{2(r)}_k $ are constants reminiscent of the ``bare'' power spectrum. The four factors in the third term determine the behavior of $ \Delta ^{2(r)}_{\mathcal R} $. Of these factors, the first two remind us of the canonical inflation. When the ratio $ p_{2*}/p_2 $ and $ c_{s*}/c_s $ both approach unity, we recover the expression for $ \Delta ^{2(r)}_{\mathcal R} $ in Ref. \cite{Urakawa:2009}. The third factor $ p_{2*}/p_2, $ is due to the added layer of complexity brought about by non-minimal coupling. The fourth and last one is due to another layer of complexity attributed to (in general,) non-constant speed of sound. 

Let us examine the behavior of the four factors above one-by-one with respect to the number of $ e $-folds $ N $ defined as $ \dd N \equiv \dd\ln a $. For the factor involving the Hubble parameter $ H $, we find from the definition $ \epsilon \equiv -\dot H/H^2 $ and that for $ N $ that 
\begin{align}
	\frac{H^2}{H_*^2} 
	&=
	e^{-2\int _{N_*}^N\dd \tilde N \, \epsilon(\tilde N) }.
\end{align}
During inflation $ \epsilon  $ goes from near zero to unity marking the end of inflation. It follows that the first factor $ H^2/H_*^2 $ exponentially decays with the number of $ e $-folds. Following a similar calculation for the second factor involving $ \tilde  \epsilon $ we find
\begin{align}
	\frac{\tilde \epsilon _*}{\tilde  \epsilon } 
	&=
	e^{-\int _{N_*}^N\dd \tilde N \, \tilde \epsilon_2(\tilde N) }.
\end{align}
Since $ \dot{\tilde \epsilon} $ has to be positive on average if $ \tilde \epsilon $ is to follow an increasing trend as inflation progresses, then $ \tilde  \epsilon _2 $ has to be positive on average as well. This implies that the factor $ \tilde \epsilon _*/\tilde \epsilon $ is also decaying with $ N $. For the third factor involving $ p_2 $, note from the definition of $ \beta _2 $ given by (\ref{beta2Define}) that 
\begin{align}
	\beta _2
	=
	\frac{\dd \ln p_2}{\dd N}.
\end{align}
In the immediately preceding section we took $ |\beta_2| $ to be much smaller than $ |\beta_1| $. Doing the same thing here implies that essentially $ p_{2*}/p_2 $ will be of the order unity for large enough $ N $ near the end of inflation. (In Ref. \cite{Alinea:2016qlf}, considering the Fakir and Unruh model of nonminimally chaotic inflation, $ p_{2*}/p_2 \sim \mathcal O(1)$.) Consequently, it cannot compete with the exponentially decaying effect of the first two factors. 

For the last factor involving the speed of sound, we note from (\ref{nearScaleInvariant}) in \textbf{Sec. \ref{barePower}} that
\begin{align}
	\delta _1(\tau ) = 2\epsilon(\tau ) - \alpha _s(\tau ).
\end{align}
Intuitively, $ \alpha _s $ is some (dimensionless) quantity measuring the deviation from exact scale invariance. For a decreasing speed of sound, the first slow-roll parameter $ \epsilon  $ should be dominant over $ \alpha _s > 0 $. However, for the limiting case where the speed of sound is constant, $ \alpha _s $ should tend to $ 2\epsilon $; the near scale invariance in this case is obeyed by virtue of the smallness of $ \epsilon $ during a significant period of inflation. 

The near scale-invariant condition above involving $ \epsilon $ and $ \delta _1 $ allows us to determine the behavior of the speed of sound with the number of $ e $-folds as
\begin{align}
	\frac{c_{s*}}{c_s}
	=
	e^{\int _{N_*}^N\dd\tilde N (2\epsilon - \alpha _s)}.
\end{align}
In combination with all other sub-results for the factors involving $ H,\, \tilde  \epsilon, $ and $p_2, $ we find
\begin{align*}
	\bigg(\frac{H^2}{H_*^2}\bigg)
	\bigg(\frac{\tilde \epsilon _*}{\tilde \epsilon } \bigg)
	\bigg(\frac{p_{2*}}{p_2} \bigg)
	\bigg(\frac{c_{s*}}{c_s}\bigg)
	=
	e^{-\int _{N_*}^N\dd \tilde N \, [
		\tilde \epsilon_2(\tilde N)  + \alpha _s(\tilde N)]
	}.
\end{align*}
implying an exponentially decaying behaviorr. In effect, the subtraction term in the equation for $ \Delta ^{2(r)}_{\mathcal R} $ given by (\ref{regPowSpecBoxed}) above becomes insignificant in the long run. In other words, the regularized power spectrum tends to the ``bare'' power spectrum with the expansion of the Universe during (and even beyond) inflation.

\section{Concluding Remarks}
\label{conclude}
The adiabatic regularization of power spectrum in canonical inflation yields a physical (regularized) power spectrum essentially the same as the ``bare'' power spectrum. In this work, we added two layers of complexities in combination namely, non-minimal coupling and varying speed of sound. Assuming a large-field inflationary scenario as in the work of Fakir and Unruh and invoking near scale invariance, we find the same behavior of the regularized power spectrum as in the work of Urakawa and Starobinsky in the canonical inflation scenario. In particular, our calculation indicates that the subtraction term is an exponentially decaying function of the number of $ e $-folds. We may see that the expansion of the Universe during (and even beyond) inflation washes out the term reminiscent of a UV regularization leading to its null effect on the power spectrum primarily constituted by frozen superhorizon modes. 

In retrospect, noting that adiabatic regularization was originally formulated to remove divergent-yielding terms in the two-point function of quantum fields in the short-wavelength limit, one may readily expect a null effect on the power spectrum. However, owing to the requirement of consistency for it to be applied to all modes and not only to UV modes, one may also expect a sort of ``tail'' of this regularization procedure extending to the long wavelength modes that could affect the power spectrum. What we have done here is a rigorous calculation aiming to shed light on this ``tail.'' As it turns out, with the help of symmetry in the form of scale invariance and some assumptions rooted in the chaotic inflationary scenario, the frozen superhorizon modes embedded in the power spectrum are not affected at all. All in all, we see this result in the most general framework considered herein as a testament to the self-consistency of adiabatic regularization and a strong support to the use of the ``bare'' power spectrum in standard calculations.

For future studies, we wish to probe deeper the connection between the null effect of adiabatic regularization on the power spectrum and scale-invariance. Physically, near scale invariance results due to an almost constant Hubble radius during inflation. If the Hubble radius is almost constant, cosmological perturbations follow the same evolutionary pathway---they exit essentially the same Hubble sphere, enhance by inflation, and then freezes. Our preliminary insight is that it might be possible that the condition of scale-invariance alone is sufficient to ensure the null effect of adiabatic regularization (or any other self-consistent regularization scheme,) on the power spectrum, independent of any model of inflation satisfying this condition. In addition to this, we may also investigate questions about the relationship between non-constant Hubble radius (say, oscillatory about a nearly constant value with respect to conformal time), symmetry in the form of scale invariance, and adiabatic regularization of the resulting power spectrum.

\begin{acknowledgments}
The seed of this study and the other connected studies published in the past few years was formed during the authors' weekly meetings with Dr. Wade Naylor and Dr. Yukari Nakanishi of the Department of Physics, Osaka University, Japan. The authors would like to acknowledge them for sharing their ideas, comments, and constructive criticisms in the early development of this work. 
\end{acknowledgments}

\end{document}